\documentclass[11pt,a4paper,twocolumn]{article}
\pdfoutput=1

\usepackage[height=8.85in,width=6.75in]{geometry}

\usepackage{graphicx,rotating}     
\usepackage[bookmarksopen,colorlinks=true,linkcolor=steelblue,
citecolor=darkred,urlcolor=darkred,linktoc=all]{hyperref}

\usepackage{amsmath,amssymb}
\usepackage{mathtools}
\usepackage{slashed}
\usepackage{bm}
\usepackage{bbm}
\usepackage{cite}
\usepackage{comment}
\usepackage[utf8]{inputenc}
\usepackage{newunicodechar}
\newunicodechar{μ}{\ensuremath{\mu}}
\usepackage{accents}
\usepackage[footnotesize]{caption}
\usepackage{chngcntr}
\counterwithin{equation}{section}
\usepackage{cancel}
\usepackage{physics}
\usepackage{here}
\usepackage{subcaption}

\usepackage[lf]{Baskervaldx} 
\usepackage[bigdelims,vvarbb]{newtxmath} 
\usepackage[cal=boondoxo]{mathalfa} 

\usepackage{stmaryrd}
\usepackage{trimclip}
\usepackage[framemethod=default]{mdframed}
\newmdenv[skipabove=7pt,
skipbelow=7pt,
rightline=false,
leftline=false,
topline=false,
bottomline=false,
backgroundcolor=gray!10,
linecolor=gray,
innerleftmargin=5pt,
innerrightmargin=5pt,
innertopmargin=5pt,
innerbottommargin=5pt,
leftmargin=0cm,
rightmargin=0cm,
linewidth=4pt]{eBox}
\newmdenv[skipabove=7pt,
skipbelow=7pt,
rightline=true,
leftline=true,
topline=true,
bottomline=true,
backgroundcolor=white,
linecolor=gray,
innerleftmargin=5pt,
innerrightmargin=5pt,
innertopmargin=5pt,
innerbottommargin=5pt,
leftmargin=0cm,
rightmargin=0cm,
linewidth=1pt]{eBox2}

\usepackage{xcolor}
\definecolor{darkred}{rgb}{0.7, 0., 0.}
\definecolor{orangered}{rgb}{1,0.27,0.}
\definecolor{steelblue}{rgb}{0.275,0.51, 0.706}
\definecolor{forestgreen}{rgb}{0.13,0.55,0.13}

\usepackage{tikz}
\usepackage{tikz-feynman}
\tikzfeynmanset{compat=1.0.0}
\pgfdeclarelayer{bg}    
\pgfsetlayers{bg,main}  
\usetikzlibrary{decorations}

\pgfdeclaredecoration{dashsoliddouble}{initial}{
  \state{initial}[width=\pgfdecoratedinputsegmentlength]{
    \pgfmathsetlengthmacro\lw{.3pt+.5\pgflinewidth}
    \begin{pgfscope}
      \pgfpathmoveto{\pgfpoint{0pt}{\lw}}%
      \pgfpathlineto{\pgfpoint{\pgfdecoratedinputsegmentlength}{\lw}}%
      \pgfmathtruncatemacro\dashnum{%
        round((\pgfdecoratedinputsegmentlength-3pt)/6pt)
      }
      \pgfmathsetmacro\dashscale{%
        \pgfdecoratedinputsegmentlength/(\dashnum*6pt + 3pt)
      }
      \pgfmathsetlengthmacro\dashunit{3pt*\dashscale}
      \pgfsetdash{{\dashunit}{\dashunit}}{0pt}
      \pgfusepath{stroke}
      \pgfsetdash{}{0pt}
      \pgfpathmoveto{\pgfpoint{0pt}{-\lw}}%
      \pgfpathlineto{\pgfpoint{\pgfdecoratedinputsegmentlength}{-\lw}}%
      \pgfusepath{stroke}
    \end{pgfscope}
  }
}
\usepackage[utf8]{inputenc}
\usepackage{textgreek}
\begin{document}
\hypersetup{pageanchor=false}
\title{\bfseries Dark Matter Interpretation of the Super-Kamiokande Antineutrino Excess in $\mathrm{U}(1)_{L_\mu-L_\tau}$ model}
\author{
Motoi Endo$^{\blacklozenge\spadesuit}$,
Yushi Mura$^{\blacklozenge}$,
Tenta Tsuji$^{\blacklozenge\spadesuit}$\\[0.5em]
{\small $^{\blacklozenge}$Theory Center, IPNS, KEK, 1-1 Oho, Tsukuba, Ibaraki 305-0801, Japan}\\
{\small $^{\spadesuit}$Graduate University for Advanced Studies (Sokendai), 1-1 Oho, Tsukuba, Ibaraki 305-0801, Japan}
}
\date{}

\makeatletter
\twocolumn[
\begin{@twocolumnfalse}
\begin{flushright}
KEK-TH-2842
\end{flushright}
\maketitle
\begin{abstract}

Recent Super-Kamiokande analyses of the diffuse supernova neutrino background, based on data across all SK phases, indicate a mild preference over the zero-DSNB hypothesis at the level of about $2.3\sigma$ with electron-like antineutrino events at $E_{\bar{\nu}_e} \simeq 20\,\mathrm{MeV}$. 
We investigate whether all or part of this excess can be accounted for by MeV-scale dark matter annihilation into neutrinos in a $\mathrm{U}(1)_{L_\mu-L_\tau}$ model.
The dark matter is a Dirac fermion with $m_\chi \simeq 22\,\mathrm{MeV}$ that annihilates via a light $Z'$ mediator into $\nu_\mu \bar{\nu}_\mu$ and $\nu_\tau \bar{\nu}_\tau$, which are partly converted into $\bar{\nu}_e$ through flavor oscillations. 
We find that this scenario can account for the excess, or a fraction thereof, while reproducing the observed relic abundance via thermal freeze-out.
We further discuss the relevant laboratory and cosmological constraints, including neutrino trident production, NA64-$\mu$, Borexino, and the contribution to $\Delta N_\mathrm{eff}$.

\end{abstract}

\vspace{1em}
\end{@twocolumnfalse}
]

\makeatother
\section{Introduction}

The origin of dark matter remains one of the central open questions in  particle physics and cosmology~\cite{Bertone:2004pz}. While most indirect searches have focused on gamma rays, charged cosmic rays, and high-energy neutrinos, MeV-scale dark matter annihilating into neutrinos offers a qualitatively different observational possibility. It can produce a low-energy, monochromatic neutrino spectrum in the energy range relevant for diffuse supernova neutrino background (DSNB) searches.

The DSNB is the integrated neutrino flux from all past core-collapse supernovae. Its prediction is subject to sizable astrophysical uncertainties, {\it e.g.}, from failed supernovae, late-time neutrino emission spectra, and the cosmic star-formation history~\cite{Horiuchi:2008jz,Nakazato:2015rya,Kresse:2020nto,Tabrizi:2020vmo}. Having operated for more than two decades with pure water, the Super-Kamiokande (SK) experiment has entered the SK-Gd phase. The gadolinium (Gd) loading improves neutron tagging efficiency and hence the sensitivity to inverse-beta-decay (IBD) signals, $\bar{\nu}_e + p \to e^+ + n$, which are the primary detection channel for DSNB searches~\cite{Beacom:2003nk,Super-Kamiokande:2021jaq,Super-Kamiokande:2025rrq}.

Combining data across all SK phases and interpreting them in terms of a DSNB component, the no-DSNB hypothesis can be rejected at the level of about $2.3\sigma$ \cite{harada_2024_12726429,Rogly:2024DSNB,Santos:2024SKGd,Beauchene:2024DSNB}. 
This mild excess may be an early indication of the DSNB, a background fluctuation, or a signal of new physics. 
In particular, although such a contribution is expected, its actual magnitude is not well determined, because the predicted DSNB normalization and spectral shape are subject to sizable astrophysical uncertainties.

Reference~\cite{Granelli:2026bem} proposed a phenomenological dark matter interpretation of this excess, considering both direct annihilation into neutrinos and annihilation via intermediate dark sector particles. 
In the former case, a monochromatic antineutrino spectrum is produced, and the global fit to all SK data yields a best-fit dark matter mass of $m_{\mathrm{DM}} = 22.1\,\mathrm{MeV}$, with a significance of $2.57\sigma$~\cite{Granelli:2026bem}.

The purpose of this paper is to realize the phenomenological interpretation of Ref.~\cite{Granelli:2026bem} within a concrete model. 
Constructing such a model is nontrivial. 
The $\mathrm{SU}(2)_L$ symmetry of the Standard Model generically ties neutrino interactions to charged-lepton interactions, and the latter are subject to stringent laboratory and cosmological constraints~\cite{Dutra:2018gmv}.

\begin{figure}[t]
\centering
\begin{tikzpicture}
\begin{feynman}
    \vertex (chi) at (-2.2,  1.0) {$\chi$};
    \vertex (chib) at (-2.2, -1.0) {$\bar\chi$};
    \vertex (v1) at (0,0);
    \vertex (v2) at (2.2,0);
    \vertex (nu) at (4.4,  1.0) {$\nu_{\mu,\tau}$};
    \vertex (nub) at (4.4, -1.0) {$\bar\nu_{\mu,\tau}$};

    \diagram*{
        (chi) -- [fermion] (v1) -- [fermion] (chib),
        (v1) -- [boson, edge label=$Z^{\prime }$] (v2),
        (v2) -- [fermion] (nu),
        (nub) -- [fermion] (v2),
    };
\end{feynman}
\end{tikzpicture}
\caption{
Dark matter annihilation into neutrinos exchanging $Z'$ boson in the $\mathrm{U}(1)_{L_\mu-L_\tau}$ model. 
The produced $\bar\nu_\mu$ and $\bar\nu_\tau$ fluxes are partially converted into $\bar\nu_e$ by flavor oscillations over Galactic distances.
}
\label{fig:process}
\end{figure}
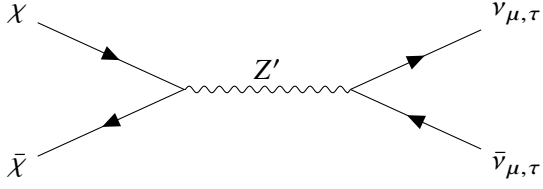

In this paper, we consider a $\mathrm{U}(1)_{L_\mu-L_\tau}$ model~\cite{Foot:1990mn,He:1990pn,He:1991qd} extended by a Dirac fermion dark matter $\chi$ charged under the symmetry. 
The model has several virtues: it is anomaly-free, its gauge boson couples to muon- and tau-flavored neutrinos but not to electrons at tree level, and MeV-scale dark matter cannot annihilate into $\mu^+\mu^-$ or $\tau^+\tau^-$ since these channels are kinematically forbidden. 
These features can make the model well suited to explaining the SK excess while avoiding the tight constraints. 
We take the dark matter mass to be
\begin{equation}
    m_\chi = 22.1\,\mathrm{MeV}.
\end{equation}
A schematic diagram of the annihilation is shown in Fig.~\ref{fig:process}. 
After flavor oscillations over Galactic distances, a fraction of the produced antineutrino flux appears as $\bar{\nu}_e$ and can contribute to the SK excess.

Light dark matter in gauged $\mathrm{U}(1)_{L_\mu-L_\tau}$ models has been studied previously in connection with thermal freeze-out and laboratory constraints~\cite{Foldenauer:2018zrz}. 
Neutrino signatures from $\mathrm{U}(1)_{L_\mu-L_\tau}$-charged dark matter at SK and Hyper-Kamiokande (HK) have been investigated in Ref.~\cite{Asai:2020qlp}, where SK data were used to place upper limits on the annihilation cross section. 
None of these studies however referred to the SK excess. 

In contrast, we examine whether a $\mathrm{U}(1)_{L_\mu-L_\tau}$ dark matter model can account for all or part of this excess, and investigate whether the preferred parameter region reproduces the observed dark matter abundance as a thermal relic and satisfies the relevant laboratory and cosmological constraints.

The rest of this paper is organized as follows. 
In Sec.~\ref{sec:model_rate}, we introduce the $\mathrm{U}(1)_{L_\mu-L_\tau}$ dark matter model, summarize the relevant interactions and decay widths, and derive the annihilation cross section, including the thermal average used for the relic-abundance calculation.
In Sec.~\ref{sec:sk_normalization}, we discuss neutrino flavor oscillations and translate the phenomenological SK excess into the annihilation cross section required in the present model.
In Sec.~\ref{sec:parameter_region}, we present the SK-motivated parameter region, compare it with the thermal relic abundance contour, and discuss the relevant laboratory and cosmological constraints, including the contribution to $\Delta N_{\rm eff}$.
Finally, Sec.~\ref{sec:conclusions} is devoted to conclusions and future prospects.

\section{Model}
\label{sec:model_rate}
We consider a gauged $\mathrm{U}(1)_{L_\mu-L_\tau}$ model extended by a Dirac fermion dark matter $\chi$ charged under the symmetry. 
The corresponding gauge boson is denoted by $Z'$.

We treat its mass $m_{Z'}$ as a free parameter without specifying the $\mathrm{U}(1)_{L_\mu-L_\tau}$ breaking sector\footnote{
    The $Z'$ mass may arise either through the Stueckelberg mechanism or through the Higgs mechanism. 
    In the latter case, we assume that the associated scalar sector is sufficiently decoupled from the low-energy phenomenology considered here.
}.

The relevant interaction Lagrangian is
\begin{equation}
\mathcal L_{\rm int}= g_\chi Z'_\rho \bar\chi\gamma^\rho\chi + g_{\mu\tau} Z'_\rho J^\rho_{L_\mu-L_\tau},
\end{equation}
where
\begin{align}
J^\rho_{L_\mu-L_\tau}
=&\,\bar L_\mu\gamma^\rho L_\mu+\bar\mu_R\gamma^\rho\mu_R-\bar L_\tau\gamma^\rho L_\tau-\bar\tau_R\gamma^\rho\tau_R .
\end{align}
Here, $L_{\mu,\tau}$ and $\mu_R, \tau_R$ denote the $\mathrm{SU}(2)_L$ lepton doublet and singlet, respectively.
Although $Z'$ does not couple to electrons at tree level, muon and tau loops induce kinetic mixing with the electromagnetic current (see, {\it e.g.,} Refs.~\cite{Araki:2017wyg, Asai:2020qlp}),
\begin{equation}
\label{eq:mixing}
\epsilon_{\rm loop}
=
\frac{e g_{\mu\tau}}{12\pi^2}
\ln\frac{m_\tau^2}{m_\mu^2}
\simeq
1.4\times10^{-2}g_{\mu\tau}.
\end{equation}
This loop suppression allows sizable annihilation into neutrinos while remaining consistent with constraints from electron couplings.
Although the same kinetic mixing also induces $Z'\to e^+e^-$, its decay width is tiny and irrelevant for the following analysis.

The decay width of the $Z'$ into the two neutrino flavors is
\begin{equation}
\Gamma_{\nu\bar\nu}=\Gamma(Z'\to\nu_\mu\bar\nu_\mu)+\Gamma(Z'\to\nu_\tau\bar\nu_\tau)=\frac{g_{\mu\tau}^2m_{Z'}}{12\pi}.
    \label{eq:gamma_nunu}
\end{equation}
If $m_{Z'}>2m_\chi$, the invisible decay into a dark matter pair is also open,
\begin{equation}
\Gamma_{\chi\bar\chi}
=
\frac{g_\chi^2 m_{Z'}}{12\pi}
\left(1+\frac{2m_\chi^2}{m_{Z'}^2}\right)
\left(1-\frac{4m_\chi^2}{m_{Z'}^2}\right)^{1/2}
\Theta(m_{Z'}-2m_\chi).
    \label{eq:gamma_chi}
\end{equation}
The total decay width of $Z'$ is given by
\begin{equation}
    \Gamma_{Z'}=\Gamma_{\nu\bar\nu}+\Gamma_{\chi\bar\chi}.
\end{equation}

We next compute the annihilation cross section. 
For two dark matter particles with equal mass, the M{\o}ller relative velocity is related to the center-of-mass energy by
\begin{equation}
v_{\rm rel}=2\left(1-\frac{4m_\chi^2}{s}\right)^{1/2},\qquad s=\frac{4m_\chi^2}{1-v_{\rm rel}^2/4}.
\label{eq:s_v}
\end{equation}
For each neutrino flavor $\alpha = \mu, \tau$, the matrix element corresponding to Fig.~\ref{fig:process} is
\begin{equation}
\begin{aligned}
    (i\mathcal M_\alpha)_{\nu \overline{\nu}}
    &=\bar v(p_2)(i g_\chi\gamma_\rho)u(p_1)\frac{-i}{s-m_{Z'}^2+i m_{Z'}\Gamma_{Z'}} \\
    &\quad\times \bar u(k_1)(i g_{\mu\tau}\gamma^\rho P_L)v(k_2),
\end{aligned}
\label{eq:amplitude}
\end{equation}
where the decay width $\Gamma_{Z'}$ is retained in the Breit--Wigner propagator to account for the $s$-channel resonance region.
After spin averaging and summing over $\nu_\mu$ and $\nu_\tau$, we obtain
\begin{equation}
    \sigma_{\nu \overline{\nu} } \, v_{\rm rel}
    =\frac{g_\chi^2g_{\mu\tau}^2(s+2m_\chi^2)}{6\pi\left[(m_{Z'}^2-s)^2+m_{Z'}^2\Gamma_{Z'}^2\right]}.
    \label{eq:sigmav_exact}
\end{equation}

In addition to the $s$-channel annihilation into neutrinos, for a light $Z'$ the process
$\chi\bar{\chi}\to Z'Z'$ through the $t$- and $u$-channel exchange of $\chi$ can be sizable.
Its annihilation cross section is given by
\begin{align}
    &\sigma_{Z'Z'} v_{\rm rel}
    =
    \frac{g_\chi^4}{4\pi s}
    \sqrt{1-\frac{4m_{Z'}^2}{s}}
    \Bigg[
    -1
    -\frac{(m_{Z'}^2+2m_\chi^2)^2}{A^2-B^2}
    \notag\\
    &+\frac{
    A^2+(m_\chi^2+m_{Z'}^2)(s-2m_\chi^2)
    }{AB}
    \ln\left(\frac{A+B}{A-B}\right)
    \Bigg]
    \Theta(s-4m_{Z'}^2),
\label{eq:sigmav_ZpZp}
\end{align}
where
\begin{align}
    A=m_{Z'}^2-\frac{s}{2},
    \qquad
    B=\frac{1}{2}
    \sqrt{(s-4m_\chi^2)(s-4m_{Z'}^2)}.
\end{align}

The thermal average is evaluated with the Gondolo--Gelmini formula~\cite{Gondolo:1990dk},
\begin{equation}
\begin{aligned}
\langle\sigma v_{\rm rel}\rangle(T)
 &=\frac{1}{8m_\chi^4 T K_2^2(m_\chi/T)}
 \int_{4m_\chi^2}^{\infty} \mathrm{d} s\,\sigma(s)(s-4m_\chi^2) \\
    &\quad \times \sqrt{s}\, K_1(\sqrt{s}/T),
\end{aligned}
\label{eq:GG}
\end{equation}
where $K_i$ is the $i$-th modified Bessel function of the second kind, and the total annihilation cross section is given by $\sigma (s) = \sigma_{\nu \overline{\nu}} (s) + \sigma_{Z' Z'} (s)$.
In the parameter scan of Sec.~\ref{sec:parameter_region}, the thermal relic curve is obtained by solving the Boltzmann equation for the dark matter abundance using this thermally averaged total annihilation cross section.

\section{Signal Normalization}
\label{sec:sk_normalization}

In the $\mathrm{U}(1)_{L_\mu-L_\tau}$ model, the primary antineutrinos produced by dark matter annihilation are $\bar\nu_\mu$ and $\bar\nu_\tau$. Since the propagation distance from the Galactic halo to SK far exceeds the oscillation length, the oscillation phases are averaged out, and the flavor-transition probability is given by~\cite{Pakvasa:2007dc}
\begin{equation}
P_{\alpha e}=\sum_i |U_{\alpha i}|^2 |U_{e i}|^2 ,
\end{equation}
where $U$ is the PMNS matrix. Assuming that $\bar\nu_\mu$ and $\bar\nu_\tau$ are produced with equal rates, the fraction of the total $\bar\nu_\mu+\bar\nu_\tau$ flux arriving at SK as $\bar\nu_e$ is
\begin{equation}
f_e\equiv \frac{P_{\mu e}+P_{\tau e}}{2}.
\end{equation}
Using the unitarity of the PMNS matrix, 
\begin{equation}
P_{\mu e}+P_{\tau e}=1-P_{ee},\qquad
P_{ee}=\sum_i |U_{e i}|^4 .
\end{equation}
Thus, $f_e$ depends only on the electron row of the PMNS matrix. Although the best-fit values of the mixing angles $\theta_{12}$ and $\theta_{13}$ differ slightly between the normal and inverted orderings, the resulting change in $f_e$ is negligible for the purposes of our analysis. Using the NuFIT-6.0 best-fit values~\cite{Esteban:2024eli}, we obtain
\begin{equation}
f_e=\frac{1-P_{ee}}{2}\simeq 0.225 .
\label{eq:fe}
\end{equation}

We recast the phenomenological fit of Ref.~\cite{Granelli:2026bem} into the $\mathrm{U}(1)_{L_\mu-L_\tau}$ model. 
In Ref.~\cite{Granelli:2026bem}, the preferred mass region is found around $17$--$25$~$\mathrm{MeV}$ at $2 \sigma$ with the best-fit value $m_{\rm DM}\simeq 22.1~{\rm MeV}$. 
The antineutrino flux is parametrized by $J_{\rm avg}\langle\sigma v\rangle_0$, where $J_{\rm avg}$ is the averaged J-factor normalized to $J_0 = 4\pi R_0 \rho_0^2$, with $R_0$ the galactocentric distance and $\rho_0$ the local dark matter density, and $\langle\sigma v\rangle_0$ denotes the present-day annihilation cross section.
For the best-fit value of $m_{\rm DM}$, it is given at $2\sigma$ by~\cite{Granelli:2026bem}
\begin{equation}
J_{\rm avg}\langle\sigma v\rangle_0= (1.8\times10^{-25}\text{--}2.3\times10^{-24}) \,{\rm cm^3\,s^{-1}}.
\label{eq:Jsigv}
\end{equation}
The approximate $2\sigma$ interval of $J_{\mathrm{avg}}$ is given by $4 \lesssim J_{\rm avg} \lesssim 17$, obtained by profiling over the generalized Navarro--Frenk--White profile parameters~\cite{Navarro:1995iw}.

In Ref.~\cite{Granelli:2026bem}, the flux normalization includes a factor $1/3$ to account for flavor equilibration by neutrino oscillations. In the present $\mathrm{U}(1)_{L_\mu-L_\tau}$ model, this factor is replaced by $f_e$ in Eq.~\eqref{eq:fe}. 
Additionally, the numerical fit results in Ref.~\cite{Granelli:2026bem} correspond to the normalization with $\kappa=2$, where $\kappa$ accounts for whether the dark-matter particle is self-conjugate. 
In our model, $\chi$ is a Dirac fermion, corresponding to $\kappa=4$. 
Hence, for a fixed antineutrino flux, the annihilation cross section required in our model is twice as large as Eq.~\eqref{eq:Jsigv}.

Therefore, the total annihilation cross section into $\bar\nu_\mu+\bar\nu_\tau$ required in our model is

\begin{equation}
    \langle\sigma v\rangle_{\rm today}
    = \frac{2}{3f_e}
    \frac{\left(J_{\rm avg}\langle\sigma v\rangle_0\right)_{\rm fit}}
    {J_{\rm avg}},
\label{eq:target_today_general}
\end{equation}
where $\left(J_{\rm avg}\langle\sigma v\rangle_0\right)_{\rm fit}$ denotes the fitted quantity in Ref.~\cite{Granelli:2026bem} and is given by Eq.~\eqref{eq:Jsigv}, while $\langle\sigma v\rangle_{\rm today}$ denotes the present-day total annihilation cross section into $\bar\nu_\mu+\bar\nu_\tau$ in our model.

In Sec.~\ref{sec:parameter_region}, we illustrate the dependence on the halo normalization by using the representative values $J_{\rm avg}=4,9,$ and $17$.
The first and last values approximately correspond to the lower and upper ends of the $2\sigma$ range inferred in Ref.~\cite{Granelli:2026bem}, while $J_{\rm avg}=9$ is shown as an intermediate benchmark.

\section{Constraints and Results}
\label{sec:parameter_region}

\begin{figure*}[t!]
    \centering

    \begin{subfigure}{0.32\textwidth}
        \centering
        \includegraphics[width=\linewidth]{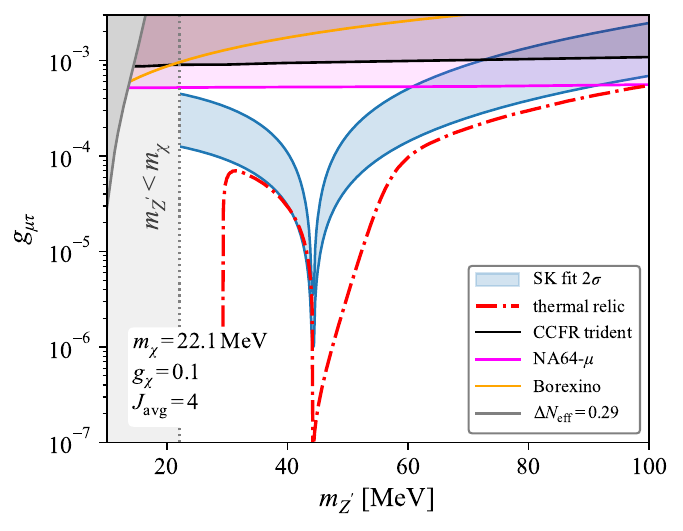}
        \caption{$J_{\rm avg}=4$}
        \label{fig:parameter_scan_J4}
    \end{subfigure}
    \hfill
    \begin{subfigure}{0.32\textwidth}
        \centering
        \includegraphics[width=\linewidth]{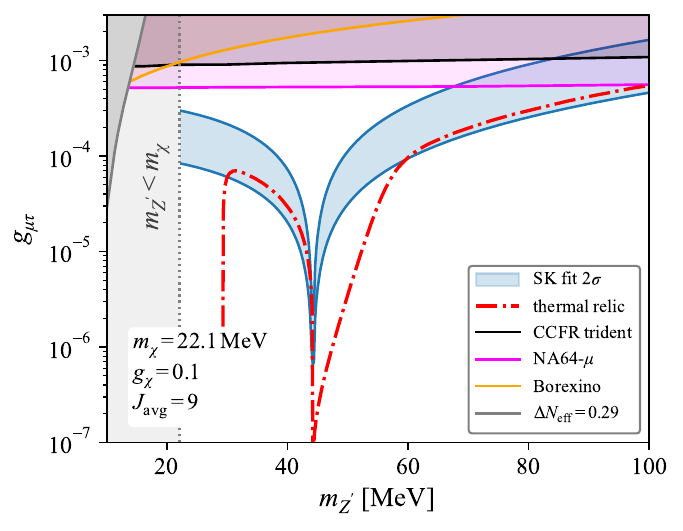}
        \caption{$J_{\rm avg}=9$}
        \label{fig:parameter_scan_J6}
    \end{subfigure}
    \hfill
    \begin{subfigure}{0.32\textwidth}
        \centering
        \includegraphics[width=\linewidth]{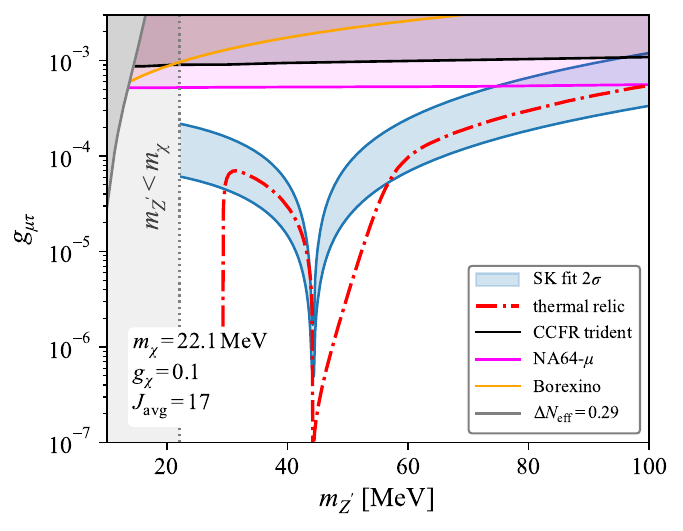}
        \caption{$J_{\rm avg}=17$}
        \label{fig:parameter_scan_J17}
    \end{subfigure}

  \caption{
Parameter region in the $(m_{Z'},g_{\mu\tau})$ plane for $m_\chi=22.1\,{\rm MeV}$ and $g_\chi=0.1$. 
The three panels correspond to the halo normalizations, $J_{\rm avg}=4$, $9$, and $17$. 
The blue shaded region shows the SK-fit $2\sigma$ region, while the red dot-dashed curve shows the parameter points reproducing the observed dark matter relic abundance. 
The black, magenta, orange, and gray shaded regions denote the constraints from CCFR trident production, NA64-$\mu$, Borexino, and $\Delta N_{\rm eff}$ from the $Z'$ decays, respectively.
The region $m_{Z'}<m_\chi$ is additionally shaded to indicate that the monochromatic SK recast based on Ref.~\cite{Granelli:2026bem} is not applied there, as discussed in the text.
}
    \label{fig:parameter_scan}
\end{figure*}

In this section, we study the parameter region of the $\mathrm{U}(1)_{L_\mu-L_\tau}$ model that is consistent with the SK-motivated annihilation cross section obtained in Sec.~\ref{sec:sk_normalization}. We first summarize the relevant laboratory and cosmological constraints, then present the numerical results comparing the SK-motivated region with the thermal relic abundance contour, and finally discuss the contribution to the effective number of neutrino species.

\subsection{Laboratory and cosmological constraints}
\label{sec:constraints}

We first summarize the constraints relevant to the parameter region of our interest.

Neutrino trident production provides an important constraint.
A light $Z'$ coupled to the $L_\mu-L_\tau$ current modifies the process
\begin{align}
\nu_\mu N\to \nu_\mu N\mu^+\mu^- ,
\end{align}
which has been measured by the CCFR experiment~\cite{CCFR:1991lpl}.
In our analysis, we follow Ref.~\cite{Altmannshofer:2014pba}.

The NA64-$\mu$ search provides another relevant constraint~\cite{NA64:2024klw}.
The signal is on-shell $Z'$ radiation from the muon, followed by invisible decays.
For $m_{Z'} < 2 m_\mu$, the invisible branching ratio is close to unity, and the NA64-$\mu$ missing-energy bound directly constrains the $Z'$ production cross section.

A sufficiently light $Z'$ can remain in thermal equilibrium with neutrinos around the time of neutrino decoupling, injecting entropy into the neutrino and thereby modifying the effective number of neutrino species $N_\mathrm{eff}$~\cite{Escudero:2019gzq}.
We adopt the 95\%~C.L. upper bound on a positive deviation from the SM prediction, $\Delta N_\mathrm{eff}\equiv N_\mathrm{eff}-N_\mathrm{eff}^{\rm SM}<0.29$, inferred from Planck~2018 combined with BAO; see Sec.~\ref{subsec:neff} for details.

This constraint becomes relevant at low mediator masses. 
The separate contribution to $N_\mathrm{eff}$ from the dark matter abundance and residual annihilation into neutrinos is discussed in Sec.~\ref{subsec:neff}.

The loop-induced kinetic mixing in Eq.~\eqref{eq:mixing} also induces a small coupling of $Z'$ to electrons, which in turn modifies neutrino--electron scattering. This is constrained by the precision solar-neutrino measurements at Borexino~\cite{Borexino:2017rsf,Amaral:2020tga}.

Before proceeding to the results, we comment on additional constraints which are not shown in the plots, as they do not affect the parameter region of interest.
Analyses of the COHERENT data~\cite{COHERENT:2020iec,COHERENT:2020ybo,COHERENT:2021xmm} in the $\mathrm{U}(1)_{L_\mu-L_\tau}$ model~\cite{AtzoriCorona:2022moj} provide a weaker constraint than those from CCFR trident production and Borexino in the mass range of our analysis.
The NA64-$e$ search~\cite{Andreev:2024lps} is most sensitive to $m_{Z'}\lesssim 2~\mathrm{MeV}$ but does not become competitive in the mass range relevant for our results.\footnote{
In the heavier-mediator region outside the mass range emphasized here, where $Z'\to\mu^+\mu^-$ is open, four-muon resonance searches at BaBar~\cite{BaBar:2016sci} and Belle~II~\cite{Belle-II:2024wtd} can become relevant.
}

Dark-matter--electron scattering through the loop-induced kinetic mixing has been studied for fermionic $\mathrm{U}(1)_{L_\mu-L_\tau}$ dark matters in Ref.~\cite{Figueroa:2024tmn}.
For $g_\chi=0.1$ used in our analysis, the current electron-recoil bounds from SENSEI-SNOLAB~\cite{SENSEI:2023zdf}, XENON1T~\cite{XENON:2019gfn}, and PandaX-4T~\cite{PandaX:2022xqx} lie above the parameter region relevant for our results.

The loop-induced coupling to electrons also allows the annihilation $\chi\bar{\chi}\to e^+e^-$, which is constrained by CMB energy injection~\cite{Slatyer:2015jla,Planck:2018vyg}.
Using Eq.~\eqref{eq:mixing}, the annihilation rates are estimated as
\begin{align}
\frac{\sigma v(\chi\bar\chi\to e^+e^-)}{\sigma v(\chi\bar\chi\to\nu\bar\nu)}
\simeq
\left(\frac{e\epsilon_{\rm loop}}{g_{\mu\tau}}\right)^2 \simeq 1.9\times10^{-5}.
\end{align}
where $\nu\bar\nu$ denotes the sum over the $\nu_\mu\bar\nu_\mu$ and $\nu_\tau\bar\nu_\tau$ final states.
Away from the immediate vicinity of the resonance, we find that the resulting Planck constraint does not exclude the thermal-relic overlap region.\footnote{
In the immediate vicinity of $m_{Z'}\simeq2m_\chi$, however, the late-time annihilation rate can be resonantly enhanced and the CMB constraint may become relevant~\cite{Drees:2021rsg}.
A precise determination of the constraint in this narrow region requires the dark-matter velocity distribution after kinetic decoupling, which can be important for Breit--Wigner resonant annihilation~\cite{Bi:2011qm}.}

Supernova observations can provide a limit on light $\mathrm{U}(1)_{L_\mu-L_\tau}$ gauge bosons.
For the minimal $\mathrm{U}(1)_{L_\mu-L_\tau}$ model in Ref.~\cite{Blinov:2025aha}, it lies below the parameter region of interest and does not exclude the phenomenologically viable region where the SK-motivated band overlaps with the thermal relic contour shown below\footnote{

When $Z'\to\chi\bar{\chi}$ is open, however, a dedicated treatment of $\chi$ transport would be required, including $\chi\nu\to\chi\nu$ scattering (see, e.g., Ref.~\cite{Manzari:2023gkt}).
We do not impose an additional model-specific supernova constraint in this region.
}.

The $\mathrm{U}(1)_{L_\mu-L_\tau}$ gauge boson gives a positive contribution to the muon anomalous magnetic moment $a_\mu$.
Following the final Fermilab measurement and in view of the unsettled situation concerning the hadronic vacuum-polarization contribution~\cite{Muong-2:2025xyk, Aliberti:2025beg}, we do not impose an additional constraint in the plot. 
Additionally, the contribution is at most $\Delta a_\mu^{Z'}=\mathcal{O}(10^{-11})$ in the SK-preferred and relic-abundance overlap region, and becomes much smaller close to the resonant region $m_{Z'} \simeq 2m_\chi$. 
It is therefore negligible compared with the present experimental and theoretical uncertainties in $a_\mu$.

\subsection{Results}

In Fig.~\ref{fig:parameter_scan}, we show the results in the $(m_{Z'}, g_{\mu\tau})$ plane for $m_\chi = 22.1\,\mathrm{MeV}$ and $g_\chi = 0.1$, for the halo normalizations $J_{\rm avg} = 4$, $9$, and $17$.\footnote{
    
    If the $L_\mu-L_\tau$ charges of the SM leptons are normalized as $Q_\mu=+1$ and $Q_\tau=-1$, one has $g_\chi=q_\chi g_{\mu\tau}$.
    Thus, $g_\chi=0.1$ and $g_{\mu\tau}\sim\mathcal{O}(10^{-4})$ correspond to $q_\chi\sim\mathcal{O}(10^3)$, with an even larger relative charge for smaller $g_{\mu\tau}$.
    We regard this hierarchy as a model-building assumption, while the physical dark-matter coupling $g_\chi=0.1$ remains perturbative.
}

The present-day annihilation cross section is evaluated at $v_{\mathrm{rel}} = 10^{-3}$.

We find that the $2\sigma$ region motivated by the SK excess (blue shaded region) overlaps with the thermal relic abundance contour (red dot-dashed curve).
In the low-$m_{Z'}$ region, $m_{Z'}\lesssim 30$\,MeV, the $\chi\bar{\chi}\to Z'Z'$ channel dominates the annihilation rate controlling thermal freeze-out.
Since this contribution is controlled predominantly by $g_\chi$, the annihilation rate eventually becomes too large to reproduce the observed relic abundance even in the limit $g_{\mu\tau} \to 0$.
As $m_{Z'}$ increases, this contribution rapidly becomes suppressed, while the thermal relic contour around the resonance, $m_{Z'}\simeq 2m_\chi$, and in the heavier-$Z'$ region is essentially determined by the $\chi\bar{\chi}\to\nu\bar{\nu}$ channel.
For $J_{\rm avg}=4$, the overlap is restricted to the resonance region, whereas for $J_{\rm avg}=9$ and $17$ overlaps are also obtained away from the resonance.

The overlap is interrupted around $m_{Z'} \gtrsim 2m_\chi$, where the decay channel $Z' \to \chi\bar{\chi}$ opens and modifies the Breit--Wigner structure of the annihilation cross section\footnote{
Even if future SK data reduce the uncertainty and shrink or remove the overlap in much of the parameter space, the $Z^\prime$ resonance region requires a separate treatment.
Near $m_{Z^\prime} \simeq 2m_\chi$, the present-day annihilation rate relevant for the SK excess and the thermally averaged rate at freeze-out are not related in a simple way, because the Breit--Wigner resonance is sampled differently at $v_{\mathrm{rel}} \simeq 0$ and at freeze-out.
Therefore, the dark-matter density-deficit problem~\cite{Leng:2026iqm} is not necessarily unavoidable in this model.
}.


The figure also displays the constraints discussed in Sec.~\ref{sec:constraints}. 
The CCFR trident bound (black) and the NA64-$\mu$ missing-energy constraint (magenta) primarily restrict the high-$m_{Z'}$ side of the SK-motivated region, while the Borexino solar-neutrino constraint (orange) and the light-$Z'$ contribution to $N_\mathrm{eff}$ (gray) become relevant at low mediator masses. 

The viable overlap region within the mass range shown is bounded by laboratory constraints from above.

In Fig.~2, we shade the region $m_{Z'}<m_\chi=22.1~\mathrm{MeV}$, where the present analysis of the SK signal is not directly applicable.
For the present-day annihilation relevant to the SK signal, the $Z'Z'$ channel is kinematically open in this region.
Since $Z'\to\chi\bar{\chi}$ is kinematically forbidden, the $Z'$ decays dominantly into neutrinos, giving a four-neutrino final state with a broad spectrum in addition to the monochromatic component.
Therefore, the monochromatic-signal fit of Ref.~\cite{Granelli:2026bem}, which we use for our main recast, cannot be directly applied to this region.
Nevertheless, as discussed above, the thermal-relic contour already disappears before reaching this region because of the efficient $Z'Z'$ annihilation.
We therefore do not perform a dedicated SK spectral fit for the shaded region.

The results shown in Fig.~\ref{fig:parameter_scan} are obtained for $g_\chi = 0.1$. 
The viable parameter region however depends on this coupling.
Away from the low-$m_{Z'}$ region where the $Z'Z'$ channel is dominant, and away from the resonance and the threshold $m_{Z'}=2m_\chi$, the annihilation cross section approximately scales as $\langle\sigma v\rangle\propto g_\chi^2g_{\mu\tau}^2$, and thus, both the SK-fit band and the thermal relic contour shift as $g_{\mu\tau} \propto 1/g_\chi$. 
As a result, the overlap region moves nearly uniformly in the vertical direction as $g_\chi$ is varied.
For instance, a larger $g_\chi$ shifts it to smaller $g_{\mu\tau}$, whereas a smaller $g_\chi$ shifts it upward, making the laboratory  constraints more relevant.

In contrast, this simple scaling does not apply near the resonance, where the Breit--Wigner denominator depends on $\Gamma_{Z'}$. In particular, once $Z' \to \chi\bar{\chi}$ opens, $\Gamma_{Z'}$ acquires a dependence on $g_\chi$, and the shape of the contours can deviate significantly from the simple $1/g_\chi$ scaling.

The results shown in Fig. 2 correspond to the case in which the SK excess is entirely attributed to the dark-matter contribution.
As a simple illustration, we suppose that dark matter accounts for only $25\%$ of the fitted excess.
The required dark-matter flux normalization is then reduced by a factor of four.
Away from the $s$-channel resonance, where the annihilation rate approximately scales as $g_{\mu\tau}^2$ for fixed $g_\chi$, this corresponds to an approximate factor-of-two reduction of the preferred $g_{\mu\tau}$.
For $J_{\rm avg}=4$, such a shift brings the high-$m_{Z'}$ SK-preferred region into overlap with the thermal-relic contour.
Thus, allowing for a nonzero DSNB contribution does not necessarily eliminate the viable thermal-relic dark-matter interpretation, but changes the preferred dark-matter normalization.
A full joint spectral fit of the DSNB and dark-matter components would depend on the assumed DSNB model and is beyond the scope of the present work.

\subsection{Dark matter contribution to $\Delta N_{\rm eff}$}
\label{subsec:neff}

MeV-scale thermal relics can modify the thermal history around neutrino decoupling. Energy or entropy injection into the photon or neutrino bath changes the neutrino-to-photon energy-density ratio, parametrized by $N_{\rm eff}$ as
\begin{equation}
\rho_\nu=N_{\rm eff}\frac{7}{8}\left(\frac{4}{11}\right)^{4/3}\rho_\gamma,\qquad \rho_\gamma=\frac{\pi^2}{15}T_\gamma^4 .
\label{eq:neff_def}
\end{equation}

In the Standard Model, non-instantaneous neutrino decoupling and finite-temperature QED effects give $N_{\rm eff}^{\rm SM}\simeq 3.044$~\cite{Mangano2005,deSalas2016,Froustey2020,Akita:2020szl,Bennett2021}. 
Current CMB observations constrain $N_{\rm eff}$ at the level of $\Delta N_\mathrm{eff} \sim 0.1$. 
For example, Planck 2018 combined with BAO gives $N_{\rm eff}=2.99\pm0.17$ at 68\% C.L.~\cite{Planck:2018vyg}. 
In the next decade, CMB-S4, in coordination with the Simons Observatory, forecasts a sensitivity to $\Delta N_{\rm eff}$ of about $0.03$ at 68\% C.L. after ten years of operation~\cite{CMB-S4:2019xqh,SimonsObservatory:2018koc}.

The dark matter mass required to explain the SK excess, $m_\chi\simeq 22.1\,{\rm MeV}$, lies above the conventional bounds on MeV-scale thermal relics from entropy transfer around neutrino decoupling~\cite{Boehm:2013jpa,Nollett:2014lwa,Escudero:2018mvt,Sabti:2019mhn}. This effect is relevant when dark matter remains thermally coupled to the neutrino bath around or after neutrino decoupling, so that the disappearance of its equilibrium population heats the neutrino bath relative to the photon bath. For example, Ref.~\cite{Sabti:2019mhn} gives, for a neutrinophilic Dirac fermion, the bound $m_\chi>11.3\,{\rm MeV}$ from Planck at 95.4\% C.L., with projected sensitivities of $15.3\,{\rm MeV}$ and $16.4\,{\rm MeV}$ for the Simons Observatory and CMB-S4, respectively.

One may also worry that residual annihilation into neutrinos after neutrino decoupling gives an additional nonthermal contribution to $N_{\rm eff}$. For a velocity-independent annihilation cross section, Ref.~\cite{Kanemura:2025byi} gives the analytic estimate
\begin{equation}
\Delta N_{\rm eff}^{\rm nth} \simeq 0.002 \left(\frac{\langle\sigma v\rangle}{10^{-24}\,{\rm cm^3\,s^{-1}}}\right)\left(\frac{1\,{\rm MeV}}{m_\mathrm{DM}}\right)\ln\left(\frac{T_{\rm dec}}{T_{\rm CMB}}\right),
\label{eq:kanemura_neff_nth}
\end{equation}
where $T_{\rm dec}=\mathcal O(0.01)\,{\rm MeV}$ is the temperature at the completion of non-instantaneous neutrino decoupling, while $T_{\rm CMB}=\mathcal O(0.1)\,{\rm eV}$ is the photon temperature at recombination. 
Taking the thermal annihilation cross section
$\langle\sigma v\rangle_{\rm th}\simeq 6\times10^{-26}\,{\rm cm^3\,s^{-1}}$ for a Dirac fermion
as a representative reference value, $m_{\rm DM}\simeq 22.1\,{\rm MeV}$, and
$\ln(T_{\rm dec}/T_{\rm CMB})\simeq 12$, we obtain
\begin{equation}
\Delta N_{\rm eff}^{\rm nth}
\simeq 6.5\times10^{-5}.
\end{equation}
This late-annihilation contribution is well below the level relevant for current $N_\mathrm{eff}$ bounds.
This estimate does not apply in the immediate vicinity of the $s$-channel resonance, where the annihilation rate can become strongly velocity dependent.
Away from this narrow resonant region, combined with the entropy-transfer effect discussed above, this suggests that neither the thermal disappearance of $\chi$ around neutrino decoupling nor the residual annihilation after freeze-out is likely to exclude the SK-motivated parameter region.

However, since the dark matter freeze-out occurs near the MeV scale, close to the epoch of neutrino decoupling, a fully reliable prediction of $N_\mathrm{eff}$ in the present model would require a dedicated calculation that consistently tracks the coupled evolution of the photon--electron bath, the neutrino bath, the dark matter energy density, and the energy density carried by the light \(Z'\), including changes in the effective number of relativistic degrees of freedom. We leave such an analysis for future work.

\section{Conclusions}
\label{sec:conclusions}

We have investigated whether the mild excess observed in the combined Super-Kamiokande analysis, interpreted in terms of the diffuse supernova neutrino background, can be accounted for, either fully or partially, by MeV-scale dark matter annihilation into neutrinos in a $\mathrm{U}(1)_{L_\mu-L_\tau}$ model.

In this scenario, a Dirac fermion dark matter with $m_\chi \simeq 22.1\,\mathrm{MeV}$ annihilates via a light $Z'$ mediator into $\nu_\mu\bar{\nu}_\mu$ and $\nu_\tau\bar{\nu}_\tau$. 
After flavor oscillations over Galactic distances, a fraction of the antineutrino flux appears as $\bar{\nu}_e$ and contributes to the SK excess.

We have shown, in the $(m_{Z'}, g_{\mu\tau})$ plane for representative choices of $J_\mathrm{avg}$, that the SK-motivated region overlaps with the thermal relic abundance contour, as shown in Fig.~\ref{fig:parameter_scan}. 
The thermal relic abundance is primarily determined by the annihilation $\chi\bar\chi\to\nu\bar\nu$ over most of the parameter region relevant for the SK excess. 
We also include the $\chi\bar\chi\to Z'Z'$ channel in the relic-abundance calculation. 
This channel dominates thermal freeze-out for $m_{Z'}\lesssim30$\,MeV, eliminating the overlap in this region, but rapidly becomes negligible toward the resonant region $m_{Z'}\simeq2m_\chi$.
The high-$m_{Z'}$ side of the overlapped region is constrained by neutrino trident production and NA64-$\mu$.

The contribution to $\Delta N_\mathrm{eff}$ from the dark matter, arising from both equilibrium entropy transfer around neutrino decoupling and residual annihilation into neutrinos after freeze-out, is expected to be small in the relevant parameter region.

Future neutrino experiments can directly test this interpretation. 
The monochromatic nature of the dark matter signal provides a way to distinguish it from the DSNB signals, whose spectrum is expected to form a broad continuum, whereas dark matter annihilation produces a localized spectral feature at $E_{\bar{\nu}_e} \simeq m_\chi$. 
In particular, owing to its excellent energy resolution, JUNO is well suited to probe neutrino-line signals in the relevant mass range~\cite{JUNO:2015zny,JUNO:2023vyz}. 
Reference~\cite{Akita:2022etk} found that a 20-year JUNO exposure can probe annihilation cross sections at the level of $4\times10^{-26}\,\mathrm{cm^3\,s^{-1}}$ for $15\,\mathrm{MeV} \lesssim m_\mathrm{DM} \lesssim 50\,\mathrm{MeV}$, covering the mass and cross-section range relevant to the present scenario. 
Hyper-Kamiokande will provide a complementary high-statistics test of the same energy window~\cite{Hyper-Kamiokande:2018ofw}
(see also Ref.~\cite{Bell:2022ycf}).

Future liquid-argon experiments such as DUNE are primarily sensitive to the $\nu_e$ flux at these energies~\cite{DUNE:2020zfm,Seadrow:2018ftp}.
They may provide a complementary test of our scenario, which predicts approximately equal $\nu_e$ and $\bar\nu_e$ fluxes~\cite{Klop:2018ltd,Buckley:2022btu}.

\section*{Acknowledgement}
This work was supported by JSPS KAKENHI Grant Numbers 22K21347 (M.E. and Y.M.) and 26KJ1238 (T.T.).
We thank Shao-Ping Li for useful comments and discussions.

\appendix
\small
\bibliographystyle{utphys}
\bibliography{ref}

\end{document}